\documentclass[useAMS,usenatbib]{mn2e}
\usepackage{graphicx}
\title[Stark broadening of helium like boron ion (B IV) spectral lines]{Stark broadening of B IV spectral lines}
\author[M. S. Dimitrijevi\'{c}, M. Christova, Z. Simi\'c, A. Kova\v cevi\'c and S. Sahal-Br\'echot]{Milan S. Dimitrijevi\'{c}$^{1,2,3,4}$\thanks{E-mail:
mdimitrijevic@aob.bg.ac.rs}, 
Magdalena Christova$^{5}$, 
Zoran Simi\'{c}$^{1}$,
Andjelka Kova\v{c}evi\'{c}$^{6}$
\newauthor
and Sylvie Sahal-Br\'{e}chot$^{3}$  \\
$^{1}$Astronomical Observatory, Volgina 7, 11060 Belgrade,
Serbia.\\ 
$^{2}$IHIS -Techno Experts, Be\v zanijski put 23, 11080 Belgrade-Zemun,
Serbia.\\ 
$^{3}$LERMA, Observatoire de Paris, PSL Research University, CNRS, Sorbonne Universities, UPMC Univ. Paris 06,\\
 5 Place Jules Janssen, 92195 Meudon Cedex, France.\\
$^{4}$Institute Isaac Newton of Chile, Yugoslavia Branch, 11060
Belgrade, Serbia.\\
$^{5}$Department of Applied Physics,  Technical University-Sofia, 1000 Sofia, Bulgaria.\\ 
$^{6}$Department for Astronomy, Faculty for Mathematics,
Studentski Trg 16, 11000 Belgrade,  Serbia.}

\begin{document}


\pagerange{\pageref{firstpage}--\pageref{lastpage}} \pubyear{2015}

\maketitle

\label{firstpage}

\begin{abstract}
 Stark broadening parameters for 157 multiplets of helium like boron (B IV) have been calculated using  the impact semiclassical 
 perturbation formalism. Obtained results have been used to investigate the regularities within spectral series.  An example
 of the influence of Stark broadening on B IV lines in DO white dwarfs is given.

\end{abstract}


\section{Introduction}

The study presents Stark broadening parameters (widths and shifts) of B IV spectral lines calculated using the impact semi-classical 
perturbation formalism \citep{b15,b16}.  B IV is a helium-like ion and such ions are rather commonly used for the diagnostic \citep{Ko86} of laboratory
\citep{TF82, Bo83, Ka83} and astrophysical plasmas \citep{Mc82}.

In astrophysics, Stark broadening data for various atomic and ionic lines (since we do 
not know {\it a priori} the chemical composition of a star) are of particular interest especially for white dwarfs, where 
this line broadening mechanism is usually the principal one \citep{Po99b,Ta03,Mi04,Si06,Hamdi08, Dimitri11,
Dufour11,Larbi12,Simic13, Simic14}. This broadening mechanism may be of interest and for the main sequence stars, especialy for
A type and late B type \citep{La88, Po99a, Po99b, Po01a, Po01b, Di03a, Di03b, Ta03, Di04, 
Mi04, Di05, Si05a, Si05b, Simic09, Simic13, Simic14}. 

We note as well the increasing astrophysical importance of Stark broadening data for various atoms and ions of trace elements,
without astrophysical meaning before   the development of satellite born telescopes which now
are providing high-resolution spectra of earlier inaccessible quality. Well-resolved
line profiles for many white dwarfs, where Stark broadening is important, have been and will
be provided for example by the Space Telescope Imaging Spectrograph (STIS), Cosmic Origins
Spectrograph (COS) and Goddard High Resolution Spectrograph (GHRS), Far Ultraviolet Spectroscopy Explorer (FUSE),
the International Ultraviolet Explorer and others.

Data on boron lines, including Stark broadening, are of interest in astrophysics but 
also for example for laboratory \citep{Bl99}, fusion \citep{Ig98} and laser produced \citep{Ni78} plasmas
investigations as well as for laser research and development \citep{Wa92}. In astrophysics, the light elements lithium, beryllium, and boron 
are of great interest for two sets of reasons, which might be categorized as cosmological and related to stellar structure \citep{Dun98}.
A general trend in nature is that the abundance of the elements versus the mass number draws a globally decreasing curve \citep{Van99, Van00}. 
However, the rare and fragile light nuclei, lithium, beryllium and boron are not generated in the normal course of stellar 
nucleosynthesis (except $^7$Li, in the galactic disk) and are, in fact, destroyed in stellar interiors. The standard Big Bang nucleosynthesis (BBN) 
theory is not effective to explain the generation of $^6$Li, $^9$Be, $^{10}$B, $^{11}$B \citep{Del93, Sch93, Tho93}, what is reflected 
in the low abundance of these simple species. Consequently the  origin and evolution of boron, are of particular interest and the corresponding
Stark broadening data are needed \citep{Ta03}.

The importance of light element abundance for the giant-branch evolution is underlined in \citet{Dun98}. The stellar structure interest steems from 
the fact that Li, Be and B undergo nuclear reactions at relatively low temperatures, approximately 2.5, 3.5, and 5 $\times 10^6$ K at densities similar 
to those in the Sun. Since these temperatures are reached not far below the convection zone and well outside the core in solar-type stars, 
circulation and destruction of the light elements can result in observable abundance changes. Observations of these changes can provide an 
invaluable probe of stellar structure and mixing. Both Li and Be abundances are greatly reduced in the giants from their initial main-sequence values. 
\citet{Dun98} report the B abundance of two giants and one dwarf in the Hyades, the latter included to evaluate explicitly the boron abundance prior 
to giant-branch evolution. They demonstrate empirically that boron contributes to the absorption spectra of cool stars. HST measurements of boron 
abundances of these objects have permitted a test of one of the basic predictions of stellar evolution theory: the growth of the convection zone as 
a star evolves up the giant branch. 

\citet{Cu99} report on the base of Hubble Space Telescope Goddard High Resolution Spectrograph spectra, the boron abundance derived for the young Orion 
solar-type member BD -  05$^o$1317. They note that, the real interstellar boron abundance and its comparison with the stellar values remains uncertain. 
Determinations of the Orion associationâs boron abundance provide unexpected results. The boron abundance derived from spectra of B-type stars 
is consistent with the expectation that the boron abundance of the Orion association should be similar to that of the solar system, but is considerably higher 
than the interstellar boron abundance for several lines of sight, including some toward Orion. A low boron abundance raises the question as to 
how the boron abundance of interstellar gas and young stars has decreased by a factor of 4 or 5 since the solar system was formed \citet{Cu99}. 

The light trace elements lithium, beryllium, and boron are at the center of astrophysical puzzles involving topics as diverse as the primordial fireball, 
interstellar (IS) or even intergalactic space, and stellar surfaces and interiors \citep{Ven02}. This role arises because boron nuclei are destroyed by 
warm protons, and thus even quite shallow mixing of the atmosphere with the interior reduces the surface abundance by bringing boron-depleted material 
to the surface. In \citet{Ven02}, a study on boron abundance of B-type stars has been presented. Boron alone is observable in hot stars where Stark
broadening is not negligible or important. A principal goal of most of these studies of hot stars was to establish the present-day boron abundance in 
order to improve our understanding of the Galactic chemical evolution of boron. Boron in hot stars, like lithium in cool stars, is shown to be a tracer 
of some of the various processes afecting a starâs surface composition that are not included in the standard models of stellar evolution. If the initial 
boron abundances of local hot stars are similar from star to star and association to association, then the large spread in boron abundances, a 
factor of at least 30 across their sample, shows that boron abundances are a clue to unraveling the nonstandard processes that afect young hot stars. 
We note that in hot stars Stark broadening is often needed for the determination of abundances and in \citet{Po99b} are analysed errors in abundances 
if it is not taken into account, especially for A-type stars.

Spectral lines of the boron ions have been observed in stellar spectra. B I lines have been observed in F and G stars \citep{Dun97}, B II in hotter stars
\citep{Cu97} and B III in early B stars \citep{Cu97, Pro99}. For example, \citet{Pro01} found B III lines in 44 early B stars.

Recently, we have calculated Stark broadening parameters for 36 B IV multiplets \citep{Dim14}. To complete as much as possible the corresponding
Stark broadening data needed in astrophysics, laboratory-, technological-, fusion-, and laser produced-plasma physics, our aim is to present in this 
work new theoretical determinations of Stark broadening parameters (full widths at half intensity and shifts) within the impact semiclassical 
perturbation approach for additional 121 helium like boron (B IV) multiplets, so that that  semiclassical perturbation Stark widths for 157 
B IV multiplets will be available. Moreover, in \citet{Dim14} data are provided only for an electron density of
$N = 10^{17}$ cm$^{-3}$, while here, Stark broadening parameters are calculated for wide range of perturber densities.
Obtained results will be used for the consideration of regulariries within spectral series, as well as for one example of the influence of Stark broadening in
comparison with Doppler broadening on B IV lines in DO white dwarfs. 

\section[]{The impact semiclassical perturbation method}

Pressure broadening of spectral lines arises when an atom, ion, or molecule which emits or absorbs light in a gas or plasma, is perturbed by 
its interactions with the other particles of the medium. Interpretation of this phenomenon is currently used for modelling of the medium and 
for spectroscopic diagnostics, since the broadening of the lines depends on the temperature and density of the medium.  The physical conditions 
in the Universe are very various, and collisional broadening with charged particles (Stark broadening) appears to be important in many domains. 
For example, at temperatures  around 10$^4$ K and densities 10$^{13}$ - 10$^{15}$ cm$^{-3}$, Stark broadening is of interest for modelling and 
analysing spectra of  A and B type stars (see e.g. \citet{La88, Po99a, Po99b, Po01a, Po01b, Ta03, Si05a, Si05b}). In white dwarfs, especially 
Stark broadening is the dominant collisional line broadening mechanism in all important layers of the atmosphere 
 \citep{Po99b,Ta03,Mi04,Si06,Dimitri11,Dufour11,Larbi12,Simic13,Simic14}. The theory of Stark broadening is well applied, especially for accurate 
 spectroscopic diagnostics and modelling. This requires the knowledge of numerous profiles, especially for trace elements, as boron in this case, 
 which are used as useful probes for modern spectroscopic diagnostics. Interpretation of the spectra of white dwarfs, which are very faint, 
 allows understanding the evolution of these very old stars, which are close to death.

 The results for Stark broadening parameters of helium like boron (B IV) multiplets have been calculated using the semi-classical perturbation formalism 
\citep{b15, b16}.We note that different later inovations and optimizations of this theoretical method have been described in details 
in \citet{b17,b18,b19,b20,b21, Sah14}. Within this theory the FWHM - Full Width at Half intensity Maximum (W) and the shift (d) of  an isolated  line, 
originating from the transition between the initial level $i$ and the final level $f$ is expressed for an ionized atom as:

\begin{eqnarray*}
W &=&N\int vf(v)dv\left( \sum\limits_{i^{\prime }\neq i}\sigma
_{ii^{\prime }}(v)+\sum\limits_{f^{\prime }\neq f}\sigma
_{ff^{\prime }}(v)+\sigma
_{el}\right) \\
\end{eqnarray*}
\begin{eqnarray}
d &=&N\int vf(v)dv\int\nolimits_{R_{3}}^{R_{D}}2\pi \rho d\rho
\sin (2\varphi _{p}).
\end{eqnarray}

\noindent where $i'$ and $f'$ are perturbing levels, $N\ $ and $\upsilon \ $ are the electron 
density and the velocity of perturbers respectively, $f(\upsilon )\ $ is the Maxwellian 
distribution of electron velocities, and $\rho \ $  the
impact parameter of the free electron colliding with the emitter.

The inelastic cross sections $\sigma _{ii^{\prime }}(\upsilon )$ (respectively $\sigma _{ff^{\prime }}(\upsilon )$),
can be expressed by an integration of the transition probability $P_{jj^{\prime
}}(\rho ,\upsilon ),\- j=i,f; \- j'=i',f'$, over the impact parameter $\rho \ $ as:

\begin{equation}
\sum_{i^{\prime }\neq i}\sigma _{ii^{\prime }}(\upsilon
)=\frac{1}{2}\pi R_{1}^{2}+\int_{R_{1}}^{R_{D}}2\pi \rho d\rho
\sum_{i^{\prime }\neq i}P_{ii^{\prime }}(\rho ,\upsilon ).
\end{equation}

\noindent and the elastic contribution to the width is given by:

\begin{eqnarray*}
\sigma _{el}=2\pi R_{2}^{2}+\int_{R_{2}}^{R_{D}}2\pi \rho d\rho
\sin ^{2}\delta+\sigma _{r},
\end{eqnarray*}
\begin{equation}
\delta =(\varphi _{p}^{2}+\varphi _{q}^{2})^{\frac{1}{2}}.
\end{equation}

Here $\sigma _{el}\ $ is the elastic cross section, while  $\varphi _{p}\ $ ($r^{-4}$) and 
 $\varphi _{q}\ $ ($r^{-3}$), are phase shifts due to the polarization and quadrupolar potential respectively, and are defined in 
 Section 3 of Chapter 2 in \citet{b15}. The cut-offs $R_{1}$,$\ R_{2}$, $\ R_{3}$, and the Debye radius $R_{D}$, as well as 
 the symmetrization procedure are described in 
 Section 1 of Chapter 3 in \citet{b16}. The contribution of Feshbach resonances, $\sigma _{r}$ is explained in \citet{b19}.
 
Within the semiclassical perturbation theory is assumed that electrons are moving along hyperbolic paths 
due to attractive Coulomb force, while in the case of ionic perturbers this force is repulsive, so that trajectories are different.     
For ion-impact broadening the formulae are analogous to Eqs. (1) - (3) for electron-impact broadening, but in the case of ion
broadening there is no term for the contribution of Feshbach resonances.

 If the considered lines are isolated, the line profile  
 $F(\omega )$ has Lorentzian form and can be expressed as:

\begin{eqnarray}
F(\omega )=\frac{W/(2\pi) }{(\omega -\omega _{if}-d)^{2}+(W/2)^{2}}
\end{eqnarray}

\noindent where

\begin{eqnarray*}
\omega _{if}=\frac{E_{i}-E_{f}}{\hbar }
\end{eqnarray*}

\noindent and $E_{i}$, $E_{f}$ are the energies of initial and final state, respectively.
Consequently, if we know  Stark width $W$ and shift $d$ it is easy to determine the profile of a considered spectral line.
  
  \section{Stark broadening parameter calculations}
  
We have calculated here, within the frame of semiclassical perturbation method \citep{b15,b16}, electron-, proton-, 
and helium ion-impact broadening parameters, full width at half maximum of intensity (FWHM - W) and shift (d)
for 157 transitions of B IV, for temperatures of 50 000K, 100 000 K, 200 000 K and 500 000K and perturber densities 
from 10$^{14}$ to 10$^{22}$ cm$^{-3}$. Energy levels needed for these calculations, have been taken from \citet{Kra08}, 
while for the needed oscillator strengths \citet{b10} method has been used, together with the tables of
\citet{Oer68} and the article of \citet{Van79} for higher levels. 

We note here, that experimental lifetimes for levels corresponding  to 385.0 \AA \- and 1170\AA \- can be found in \citet{Ke75}.
 However, for semiclassical perturbation calculations of Stark broadening parameters we need a consistent set  of oscillator
strengths and if  we use  a mixture of oscillator strengths obtained with various methods this  may result in worse
agreement with experimental data for Stark broadening,
 as  shown in \citet{Di94}, since the corresponding summation rule for oscillator strengts for transitions starting or
ending on a particular energy level, may be violated.

The complete results for all 157 transitions, are provided in electronic, computer readable form, as additional data
in the on line journal editionas as Tables S1-S7. In Table S1 are results for 10$^{14}$ cm$^{-3}$ and in
Tables S2-S7 for 10$^{17}$ to 10$^{22}$ cm$^{-3}$. As an example, here are given in Table 1,  the results 
for two transitions as a sample in order to demonstrate the form of additional data.  

 It should be noticed that wavelengths in Table 1 as well as in Tables S1-S7 differs from experimental ones since they
 are calculated. Since the Stark broadening parameter calculation depends only on relative positions of energy
levels they do not depend on absolute positions of energy levels when expressed in angular frequency units. 
The  Stark width in \AA  \- \-may be expressed in angular frequency units using the formula:

\begin{equation}
W(\mathring{A})=\frac{\lambda ^{2}}{2\pi c}W(s^{-1})
\end{equation}

\noindent where $c$ is the speed of light.
  If a correction of the width or shift for the difference
between calculated and experimental wavelength is necessary this can be made for the width, and similarly for the shift as:

\begin{equation}
W_{cor}=\left( \frac{\lambda _{\exp }}{\lambda }\right) ^{2}W.
\end{equation}

\noindent Here, with $W$$_{cor}$ is denoted the corrected width,
$\lambda$$_{exp}$ is the experimental, $\lambda$ the
calculated wavelength and $W$ the width from Table 1, or Tables S1-S7.
 
A parameter \emph{C} \citep{DS84}, provided in Table 1 and Tables 1S-7S, enables to determine the maximal perturber density 
when the line may be considered as isolated, by division of  \emph{C} with the corresponding
full width at half maximum. The validity of impact approximation is checked for all results estimating the value of \emph{N}\emph{V},
where  \emph{V} is the collision volume and \emph{N} the perturber density. If \emph{N}\emph{V}$<$ 0.1, we assume that 
the impact approximation is valid \citep{b15,b16}. If \emph{N}\emph{V}$>$ 0.5 the corresponding results for width and shift  are not 
shown in tables since the impact approximation breaks down, while for  0.1$<$\emph{N}\emph{V}$\leq$0.5 befor the corresponding 
Stark broadening parameters is an asterisk to notice that this result is on the limit of validity of impact  approximation.
In the cases when, for the given (\emph{T} and \emph{N}), the impact approximation is not valid, for the estimation of ion broadening 
is convenient quasistatic approach \citep{Griem74,b18}. If neither impact nor quasistatic approximation are valid, one can use a unified-type 
theory, as for example the approach of \citet{Barnard74} providing a simple analytical formula, convenient to use for such a
case.

\begin{table*}
\caption{This table gives electron-, proton-, and helium ion-impact 
broadening parameters for B IV lines. Calculated wavelength of the transitions (in \AA)
and parameter \emph{C} are also given. This parameter, when divided
with the corresponding Stark width, gives an estimate for the
maximal pertuber density for which the line may be treated as
isolated. The Stark broadening parameters for 157 B IV lines are available
in its entirety in machine-readable
form in the online journal as additional data. Results for perturber 
density of 10$^{14}$ cm$^{-3}$ are in Table 1, and for  10$^{17}$ cm$^{-3}$ - 
10$^{22}$ cm$^{-3}$ in Tables 2-7. In all Tables temperatures are from 20 000
to 500 000 K. A positive shift is towards the red part of the spectrum. Data for 36 multiplet for electron density of  10$^{17}$ cm$^{-3}$
are from \citet{Dim14} and here are added in order to complete Stark broadening data for B IV.
A portion, for a
perturber density of 10$^{17}$ cm$^{-3}$  is shown
here for guidance regarding its form and content.}
\label{tab:2}       
\begin{center}
\begin{tabular}{crrrrrrrr}
\hline  PERTURBER DENSITY =& 1E+17cm-3   & & & & \\

   &  & ELECTRONS& &PROTONS& &IONIZED& HELIUM\\
TRANSITION  &  T(K)&  WIDTH(A)&  SHIFT(A) & WIDTH(A)&  SHIFT(A) &  WIDTH(A)&  SHIFT(A)\\
\hline
SINGLETS & & & & & & \\

  B IV 2s-2p&  20000.& 0.171   & -0.930E-02& 0.354E-03&-0.208E-02& 0.565E-03&-0.204E-02\\
  4492.1 A  &  50000.& 0.110   & -0.646E-02& 0.206E-02&-0.468E-02& 0.252E-02&-0.437E-02\\
 C= 0.45E+21& 100000.& 0.818E-01&-0.696E-02& 0.466E-02&-0.692E-02& 0.476E-02&-0.617E-02\\
            & 200000.& 0.623E-01&-0.687E-02& 0.835E-02&-0.924E-02& 0.744E-02&-0.774E-02\\
            & 300000.& 0.538E-01&-0.673E-02& 0.101E-01&-0.103E-01& 0.878E-02&-0.864E-02\\
            & 500000.& 0.454E-01&-0.644E-02& 0.122E-01&-0.117E-01& 0.105E-01&-0.979E-02\\
             
  & & & & & & & & \\
  
  B IV 2s-3p & 20000.& 0.336E-02&-0.199E-03& 0.137E-03&-0.230E-03& 0.151E-03&-0.201E-03\\
   380.9 A   & 50000.& 0.233E-02&-0.189E-03& 0.319E-03&-0.375E-03& 0.303E-03&-0.316E-03\\
 C= 0.20E+18 &100000.& 0.181E-02&-0.181E-03& 0.471E-03&-0.457E-03& 0.402E-03&-0.386E-03\\
             &200000.& 0.143E-02&-0.163E-03& 0.626E-03&-0.546E-03& 0.513E-03&-0.462E-03\\
             &300000.& 0.126E-02&-0.145E-03& 0.765E-03&-0.598E-03& 0.573E-03&-0.499E-03\\
             &500000.& 0.107E-02&-0.125E-03& 0.909E-03&-0.655E-03& 0.675E-03&-0.552E-03\\
            
\hline
\end{tabular}
\end{center}
\end{table*}

The dependence of the broadening parameters of spectral lines due to impacts with charged particles versus principal quantum number within a spectral series 
is a useful information. If we know the trend of Stark broadening parameters within a spectral series, it is possible to interpolate or extrapolate 
the eventually missing values within the considered series. The regular behavior of electron Stark broadening width versus principal quantum number within 
three  spectral series, 1s$^2$ $^1$S - 1snp $^1$P$^Â°$, 1s2s $^1$S - 1snp $^1$P$^Â°$ and 1s3p $^1$P$^Â°$ - 1snd $^1$D is presented in Fig. 1. The widths 
increase with principal quantum number which reflects the decrease of the distance to the nearest perturbing levels with the increase of the principal 
quantum number.      

\begin{figure}
\centerline{\includegraphics[width=\columnwidth,
height=0.75\columnwidth]{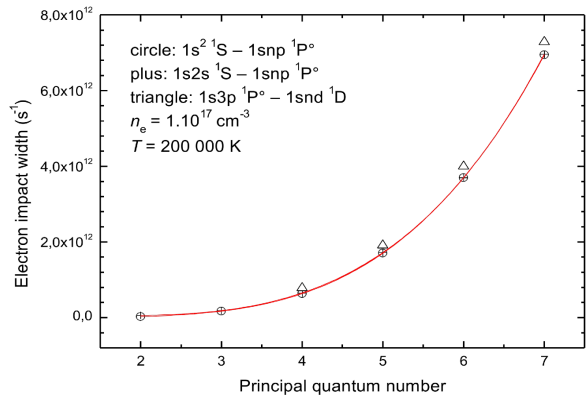}} \caption{Principal quantum number dependence of the electron-impact width (in angular frequency units) for multiplets within 
1s$^2$ $^1$S - 1snp $^1$P$^{\rm o}$, 1s2s $^1$S - 1snp $^1$P$^{\rm o}$ and 1s3p $^1$P$^{\rm o}$ - 1snd $^1$D spectral series at an electron density of
1.10$^{17}$ cm$^{-3}$ and T = 200 000 K. }
\label{fig:Qtr1}
\end{figure}

The calculated data for 1s$^2$ $^1$S - 1snp $^1$P$^Â°$ and 1s2s $^1$S - 1snp $^1$P$^Â°$  spectral series are fitted with polynomial 
function of 4-th order; the fit-curve is added in the figure. The data for 1s3p $^1$P$^Â°$ - 1snd $^1$D spectral series are fitted with polynomial function 
of 3-th order. The polynomial function permits the interpolation and extrapolation to obtain electron impact width for other lines 
within the series where there is a lack of the data to perform the calculations. The analytical function up to 4-th order is given below: 

\begin{equation}
W(n)= a_4n^4 + a_3n^3 + a_2n^2 + a_1n + a_0
\end{equation}

\noindent where W(n) is the full width at half-maximum in angular frequency units. The corresponding coefficients for the three series are given in Table 2.

\begin{table*}
\caption{Coefficients of the fitting polynomial functions for three spectral series. The first column presents the corresponding transition,
the next 5 columns give the values of the coefficients in the fitting polynomial functions.}
\label{tab:3}       
\begin{center}
\begin{tabular}{crrrrrr}
\hline Transition  &a$_0$ & a$_1$& a$_2$& a$_3$&a$_4$\\
\hline
 
 1s$^2$ $^1$S - 1snp $^1$P$^{\rm o}$& -4.2421E+11&5.2977E+11&-2.33982E+11&3-8929E+10&7.37826E+8\\
\hline
1s2s $^1$S - 1snp $^1$P$^{\rm o}$&1.06075E+11&-1.54335E+9&-3.69292E+10&7.9373E+9&2.48152E+9\\
\hline
1s3p $^1$P$^{\rm o}$ - 1snd $^1$D&1.00003E+12&-1.91806E+11&-1.26934E+11&4.03773E+10&\\

\hline
\end{tabular}
\end{center}
\end{table*}

As an example of astrophysical significance of obtained results Stark and Doppler line widths for  B IV 2s $^1$S - 2p $^1$P$^{\rm o}$
($\lambda$ = 4492.1 \AA) spectral line are compared in Fig. 2, as a function of Rosseland optical depth ($\tau$) for an atmospheric model \citep{Wes81}  
of helium rich DO white dwarf with the effective temperatures  T$_{eff}$ = 80 0000 K, and log $g$ = 8. 
One can see that Stark broadening dominates in the large part of the considered atmosphere.  

\begin{figure}
\centerline{\includegraphics[width=\columnwidth,
height=0.75\columnwidth]{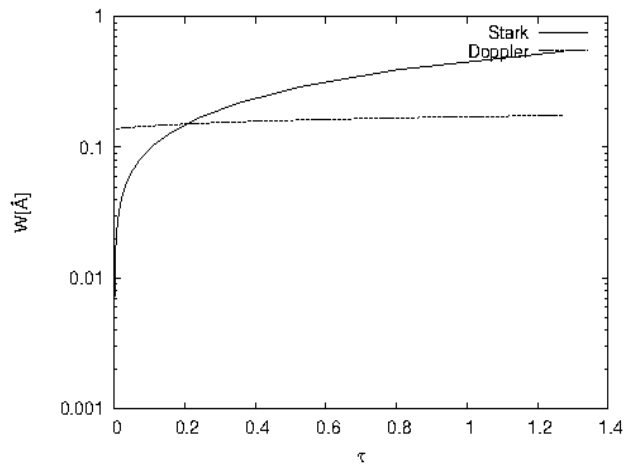}} \caption{Stark and Doppler widths for B IV 2s $^1$S - 2p $^1$P$^{\rm o}$
($\lambda$ = 4492.1 \AA) spectral line as a function of Rosseland optical depth ($\tau$). Stark  and Doppler widths 
 are shown for an atmospheric model \citep{Wes81}  with effective temperatures  T$_{eff}$ = 80 0000 K, and log $g$ = 8.}
\label{fig:Qtr2}
\end{figure}

\section{Conclusions}
New Stark broadening parameters for 157 multiplets of B IV have been calculated within the frame of the semi-classical perturbation formalism. 
The results for Stark broadening parameters of boron lines could be applicable for the adequate interpretation of the observed spectra in 
astrophysics, astrochemistry, and cosmology, in technological plasma research, for thermonuclear reaction devices, and for laser produced plasma 
investigation. We also note that the Stark broadening data obtained in the present research, will be inserted in the STARK-B database 
\citep{ssb12, starkb, Sah10, Sah15}, which is a part of Virtual Atomic and Molecular Data Center (VAMDC - \citep{vamdc,Rixon11}.
Additionally, it has been confirmed that the behaviour of Stark broadening parameters within B IV spectral series is regular, enabling 
the interpolation and extrapolation of new data.

Also, an example of the significance of Stark broadening
mechanism for BIV lines in DO white dwarf atmospheres
is given.

\section*{Acknowledgments}

This work is  a part of the project 176002 "Influence of
collisional processes on astrophysical plasma line shapes"
supported by the Ministry of Education, Science and Technological
Development of Serbia. It is also the result of  Short term Scientific Mission CM0805 within COST programme on
The Chemical Cosmos: Understanding Chemistry in Astronomical Environments. Partial financial support from project No
6572-20 Technical University - Sofia within Euroatom Programme is also acknowledged.
This work has been supported as well, by the Paris Observatory,
the CNRS and the PNPS (Programme National de Physique Stellaire, INSU-CNRS).

\bsp \label{lastpage}


\begin{thebibliography}{99}

\bibitem[\protect\citeauthoryear{Barnard, Cooper \& Smith}{1974}]{Barnard74} Barnard A. J., Cooper J., Smith E. W., 1974, 
J. Quant. Spectrosc. Radiat. Transfer, 14,
1025

\bibitem[\protect\citeauthoryear{Bates \& Damgaard}{1949}]{b10} Bates D. R., Damgaard A.,
1949, Philos. Trans. R. Soc. London A, 242, 101

\bibitem[\protect\citeauthoryear{Blagojevi\'c et al.}{1999}]{Bl99} Blagojevi\'c, B., Popovi\'c, M. V., Konjevi\'c, N., Dimitrijevi\'c, M. S., 1999, 
    J. Quant. Spectrosc. Radiat. Transfer, 61, 361 
    
\bibitem[\protect\citeauthoryear{Boiko et al.}{1983}]{Bo83} Boiko, V. A, Faenov, A. Ya., Hahalin, S. Ya., Pikuz, S. A., Shilov, K. A., Skobelev, I. Yu., 
1983, J. Phys. B, 16, L77   

\bibitem[\protect\citeauthoryear{Cunha, Smith \& Lambert}{1999}]{Cu99} Cunha, K., Smith, V. V., Lambert, D. L., 1999, ApJ, 519, 844

\bibitem[\protect\citeauthoryear{Cunha et al.}{1997}]{Cu97} Cunha, K., Lambert, D. L., Lemke, M., Gies, D. R., Roberts, L. C., 1997, ApJ, 478, 211

\bibitem[\protect\citeauthoryear{Delbourgo-Salvador \& Vangioni-Flam}{1993}]{Del93} Delbourgo-Salvador, P., Vangioni-Flam, E., 1993, 
in: Prantzos et al. (Eds.), Origin and Evolution of Elements,
Cambridge University Press, Cambridge, p. 52

\bibitem[\protect\citeauthoryear{Dimitrijevi\'{c} \& Sahal-Br\'{e}chot}{1984}]{DS84} Dimitrijevi\'{c} M. S., Sahal-Br\'{e}chot
S., 1984, J. Quant. Spectrosc. Radiat. Transfer, 31, 301

\bibitem[\protect\citeauthoryear{Dimitrijevi\'{c} \& Sahal-Br\'{e}chot}{1994}]{Di94} Dimitrijevi\'{c} M. S., Sahal-Br\'{e}chot
S., 1994, Phys. Scr., 49, 661

\bibitem[\protect\citeauthoryear{Dimitrijevi\'{c} \& Sahal-Br\'{e}chot}{1996}]{b21} Dimitrijevi\'{c} M. S., Sahal-Br\'{e}chot
S., 1996, Phys. Scr., 54, 50

\bibitem[\protect\citeauthoryear{Dimitrijevi\'{c}, Sahal-Br\'{e}chot \& Bommier}{1991}]{b20} Dimitrijevi\'{c} M. S., Sahal-Br\'{e}chot S., 
Bommier V., 1991, A\&AS, 89,
581

\bibitem[\protect\citeauthoryear{Dimitrijevi\'c et al.}{2003a}]{Di03a} Dimitrijevi\'c M. S., Jovanovi\' c P., Simi\' c, Z., 2003a, A\&A, 410, 735

\bibitem[\protect\citeauthoryear{Dimitrijevi\'c et al.}{2003b}]{Di03b} Dimitrijevi\'c M. S., Ryabchikova T., Popovi\'c L. \v C., 
Shulyak D., Tsymbal, V., 2003b, A\&A, 404, 1099

\bibitem[\protect\citeauthoryear{Dimitrijevi\'c et al.}{2004}]{Di04} Dimitrijevi\'c M. S., Da\v ci\' c M., Cvetkovi\' c Z., Simi\' c Z., 
2004, A\&A, 425, 1147

\bibitem[\protect\citeauthoryear{Dimitrijevi\'c et al.}{2005}]{Di05} Dimitrijevi\'c M. S., Ryabchikova T., Popovi\'c L. \v C., Shulyak D., 
Khan S., 2005, A\&A, 435, 1191

\bibitem[\protect\citeauthoryear{Dimitrijevi\'{c} et al.}{2011}]{Dimitri11} Dimitrijevi\'{c} M. S., Kova\v{c}evi\'{c} A., Simi\'{c} Z., 
Sahal-Br\'{e}chot S., 2011, Baltic Astronomy, 20, 580

\bibitem[\protect\citeauthoryear{Dimitrijevi\'c et al.}{2014}]{Dim14} Dimitrijevic, M. S., Christova, M., Simi\'c, Z., Kova\v cevi\'c, A., Sahal-Br\'echot, S., 
2014, Adv. Space Res., 54, 1195

\bibitem[\protect\citeauthoryear{Dubernet et al.}{2010}]{vamdc} Dubernet M. L., Boudon V., Culhane J. L., Dimitrijevi\'c M. S., 
Fazliev A. Z. et al., 2010, J. Quant. Spectrosc. Radiat. Transfer, 111, 2151, http://www.vamdc.org

\bibitem[\protect\citeauthoryear{Dufour et al.}{2011}]{Dufour11} Dufour P., Ben Nessib
N., Sahal-Br\'{e}chot S., Dimitrijevi\'{c} M. S., 2011, Baltic
Astronomy, 20, 511

\bibitem[\protect\citeauthoryear{Duncan et al.}{1998}]{Dun98} Duncan D. K., Peterson R. C., Thorburn J. A., Pinsonneault M. H., 1998, {\bf ApJ}, 499, 871

\bibitem[\protect\citeauthoryear{Duncan et al.}{1997}]{Dun97} Duncan, D. K., Primas, F., Rebull, L. M., Boesgaard, A. M., Deliyannis, Constantine P.,
Hobbs, L. M., King, J. R., Ryan, S. G., 1997, ApJ, 488, 338

\bibitem[\protect\citeauthoryear{Fleurier, Sahal-Br\'{e}chot \& Chapelle}{1977}]{b19} Fleurier C., Sahal-Br\'{e}chot S., Chapelle J., 1977, J. Quant.
Spectrosc. Radiat. Transfer, 17, 595

\bibitem[\protect\citeauthoryear{Griem}{1974}]{Griem74} Griem H. R.,
1974, Spectral line Broadening by Plasmas. McGraw-Hill, New York

\bibitem[\protect\citeauthoryear{Hamdi et al.}{2008}]{Hamdi08} Hamdi R., Ben Nessib N., Milovanovi\'{c} N., Popovi\'{c} L. \v{C}., Dimitrijevi\'{c}
M. S., Sahal-Br\'{e}chot S., 2008, MNRAS, 387, 871

\bibitem[\protect\citeauthoryear{Iglesias et al.}{1998}]{Ig98} Iglesias, E., Griem, H., Welch, B., Weaver, J., 1998, 
    Astrophys. Space Sci., 256, 327 
    
\bibitem[\protect\citeauthoryear{K\"allne, K\"allne \& Pradhan}{1983}]{Ka83} K\"allne, E., K\"allne, J., Pradhan, A. K., 1983, Phys, Rev. A, 22, 1476

\bibitem[\protect\citeauthoryear{Kernahan et al.}{1975}]{Ke75} Kernahan, J. A., Pinnington, E. H., Livingston, A. E., Irwin, D. J. G., 1975, Phys. Scr., 12, 319

\bibitem[\protect\citeauthoryear{Kolk, K\"onig \& Kunze}{1986}]{Ko86} Kolk, K.-H., K\"onig, R., Kunze, H. -J., 1986, Phys. Rev. A, 33, 747

\bibitem[\protect\citeauthoryear{Kramida et al.}{2008}]{Kra08} Kramida, A. E., Ryabtsev, A. N., Ekberg, J. O., Kink, I., Mannervik, S., Martinson, I.,  
2008, Phys. Scr., 78, 025302

\bibitem[\protect\citeauthoryear{Lanz et al.}{1988}]{La88} Lanz T., Dimitrijevi\'c M. S., Artru M. C., 1988, A\&A, 192, 249

\bibitem[\protect\citeauthoryear{Larbi-Terzi et al.}{2012}]{Larbi12} Larbi-Terzi N., Sahal-Br\'{e}chot S.,
Ben Nessib N., Dimitrijevi\'{c} M. S., 2012, MNRAS, 423, 766

\bibitem[\protect\citeauthoryear{McKenzie \& Landecker}{1982}]{Mc82} McKenzie, D. L. \& Landecker, P. B., 1982, ApJ, 259, 372

\bibitem[\protect\citeauthoryear{Milovanovi\' c et al.}{2004}]{Mi04} Milovanovi\' c N., Dimitrijevi\'c M. S., Popovi\'c L. \v C., 
Simi\' c, Z., 2004, A\&A, 417, 375

\bibitem[\protect\citeauthoryear{Nicolosi et al.}{1978}]{Ni78} Nicolosi, P., Garifo, L., Jannitti, E., Malvezzi, A. M., Tondello, G., 1978,
    Nuovo Cimento B, 48, 133 

\bibitem[\protect\citeauthoryear{Oertel \& Shomo}{1968}]{Oer68} Oertel G. K., Shomo L. P., 1968, ApJS, 16, 175
 
\bibitem[\protect\citeauthoryear{Popovi\'c et al}{1999a}]{Po99a} Popovi\'c L. \v C., Dimitrijevi\'c M. S., Ryabchikova T., 1999a, A\&A, 350, 719

\bibitem[\protect\citeauthoryear{Popovi\'c et al.}{1999b}]{Po99b} Popovi\'c L. \v C., Dimitrijevi\'c M. S., Tankosi\' c D., 1999b, A{\&}AS, 139, 617

\bibitem[\protect\citeauthoryear{Popovi\'c et al}{2001a}]{Po01a} Popovi\'c L. \v C., Milovanovi\' c N., Dimitrijevi\'c M. S., 2001a, A\&A, 365, 656

\bibitem[\protect\citeauthoryear{Popovi\'c et al}{2001b}]{Po01b} Popovi\'c L. \v C., Simi\' c S., Milovanovi\' c N., Dimitrijevi\'c M. S., 
2001b, ApJS, 135, 109

\bibitem[\protect\citeauthoryear{Proffitt \& Quigley}{2001}]{Pro01} Proffitt, C. R., Quigley, M. F., 2001, ApJ, 548, 429

\bibitem[\protect\citeauthoryear{Proffitt et al.}{1999}]{Pro99} Proffitt, C. R., J\"onsson, P., Litz\'en, U., Pickering, J. C., Wahlgren, G. M.,
1999, ApJ, 516, 342

\bibitem[\protect\citeauthoryear{Rixon et al.}{2011}]{Rixon11} Rixon G., Dubernet M. L., Piskunov, N., Walton, N., Mason, N., Le Sidaner, P., 
Schlemmer, S., Tennyson, J., et al.,
2011 7$^{th}$ International Conference on Atomic and Molecular Data and their Applications -ICAMDATA-2010, AIP Conf. Proc. 1344, 107

\bibitem[\protect\citeauthoryear{Sahal-Br\'{e}chot}{1969a}]{b15} Sahal-Br\'{e}chot S., 1969a,
A\&A, 1, 91

\bibitem[\protect\citeauthoryear{Sahal-Br\'{e}chot}{1969b}]{b16} Sahal-Br\'{e}chot S., 1969b,
A\&A, 2, 322

\bibitem[\protect\citeauthoryear{Sahal-Br\'{e}chot}{1974}]{b17} Sahal-Br\'{e}chot S., 1974,
A\&A, 35, 319

\bibitem[\protect\citeauthoryear{Sahal-Br\'{e}chot}{1991}]{b18} Sahal-Br\'{e}chot S., 1991,
A\&A, 245, 322

\bibitem[\protect\citeauthoryear{Sahal-Br\'{e}chot et al.}{2010}]{Sah10} Sahal-Brechot, S., J. Phys.: Conf. Ser., 257, 012028

\bibitem[\protect\citeauthoryear{Sahal-Br\'{e}chot et al.}{2014}]{Sah14}  Sahal-Br\'echot S., Dimitrijevi\'c M. S., Ben Nessib N., 2014,  Atoms, 2, 225 

\bibitem[\protect\citeauthoryear{Sahal-Br\'echot et al.}{2012}]{ssb12}
Sahal-Br\'echot, S., Dimitrijevi\'c, M.S., Moreau N., 2012, J. Phys.: Conf. Ser., 397, 012019

\bibitem[\protect\citeauthoryear{Sahal-Br\'{e}chot et al.}{2015a}]{starkb}Sahal-Br\'echot S., Dimitrijevi\'c M. S., 
Moreau N. 2015a, STARK-B database, [online]. 
Available: http://stark-b.obspm.fr [June 19, 2015]. Observatory of Paris, LERMA and Astronomical Observatory of Belgrade

\bibitem[\protect\citeauthoryear{Sahal-Br\'{e}chot et al.}{2015b}]{Sah15} Sahal-Br\'echot S., Dimitrijevi\'c M. S., Moreau N., Ben Nessib N., 2015b,  
Phys. Scripta, 50, 054008

\bibitem[\protect\citeauthoryear{Schramm}{1993}]{Sch93} Schramm, D. N., 1993, in: Prantzos et al. (Eds.), Origin and Evolution of the Elements, 
Cambridge University Press,
Cambridge, p. 112

\bibitem[\protect\citeauthoryear{Shore \& Menzel}{1965}]{Shore65} Shore B. W., Menzel D., 1965, ApJS, 12,
187

\bibitem[\protect\citeauthoryear{Simi\'{c} et al.}{2005a}]{Si05a} Simi\'{c} Z., Dimitrijevi\'{c} M. S., 
Popovi\'{c} L. \v C., Da\v ci\'c M., 2005a, J. Appl. Spectrosc., 72, 443

\bibitem[\protect\citeauthoryear{Simi\'{c} et al.}{2005b}]{Si05b} Simi\'{c} Z., Dimitrijevi\'{c} M. S., Milovanovi\'c N., 
Sahal-Br\'echot S., 2005b, A\&A, 441, 391

\bibitem[\protect\citeauthoryear{Simi\'{c} et al.}{2006}]{Si06} Simi\'{c} Z., Dimitrijevi\'{c} M. S., Popovi\'{c} L. \v C., 
Da\v ci\'c M., 2006, New Astron., 12, 187

\bibitem[\protect\citeauthoryear{Simi\'{c}, Dimitrijevi\'{c} \& Kova\v{c}evi\'{c}}{2009}]{Simic09} Simi\'{c} Z., Dimitrijevi\'{c} M. S., 
Kova\v{c}evi\'{c} A., 2009, New Astron. Rev., 53, 246

\bibitem[\protect\citeauthoryear{Simi\'{c}, Dimitrijevi\'{c} \& Sahal-Br\'echot}{2013}]{Simic13} Simi\'{c} Z., Dimitrijevi\'{c} M. S., 
Sahal-Br\'echot S., 2013, MNRAS, 432, 2247

\bibitem[\protect\citeauthoryear{Simi\'{c}, Dimitrijevi\'{c} \& Popovi\'{c}}{2014}]{Simic14} Simi\'{c} Z., Dimitrijevi\'{c} M. S., 
Popovi\'{c} L. \v C., 2014, Adv. Space Res., 54, 1231

\bibitem[\protect\citeauthoryear{Tankosi\'c, Popovi\'c \& Dimitrijevi\'c}{2003}]{Ta03} Tankosi\'{c} D., Popovi\'c L. \v C., 
Dimitrijevi\'{c} M. S., 2003, A\&A, 399, 795

\bibitem[\protect\citeauthoryear{TFR Group, Doyle \& Schwob,}{1982}]{TF82} TFR Group, Doyle, J. G., Schwob, J. L., 1982, J. Phys. B, 15, 813

\bibitem[\protect\citeauthoryear{Thomas et al.}{1993}]{Tho93} Thomas, D., Schramm, D. N., Olive, K. A., Fields, B. D., 1993, ApJ, 406, 569

\bibitem[\protect\citeauthoryear{Van Regemorter, Hoang Binh \& Prud'homme}{1979}]{Van79} Van Regemorter H., Hoang Binh Dy, Prud'homme
M., 1979, J. Phys. B, 12, 1073

\bibitem[\protect\citeauthoryear{Vangioni-Flam \& Cass\'e}{1999}]{Van99} Vangioni-Flam E., Cass\'e M., 1999, Astrophys. Space Sci., 265, 77 

\bibitem[\protect\citeauthoryear{Vangioni-Flam, Cass\'e \& Audouze}{2000}]{Van00} Vangioni-Flam E., CassÃ© M., Audouze J., 2000, Phys. Rep. 333-334, 365

\bibitem[\protect\citeauthoryear{Venn et al.}{2002}]{Ven02} Venn K. A., Brooks A. M., Lambert, D. L., Lemke, M., Langer, N., Lennon, D. J., 
Keenan, F. P., 2002, ApJ, 565, 571 

\bibitem[\protect\citeauthoryear{Wang et al.}{1992}]{Wa92} Wang, J. S., Griem, H. R., Huang, Y. W., B\"ottcher, F., 1992, 
    Phys. Rev. A, 45, 4010    

\bibitem[\protect\citeauthoryear{Wesemael}{1981}]{Wes81} Wesemael F., 1981, ApJS, 4{\bf 5, 177}

\end{thebibliography}
\end{document}